\documentclass[graybox]{svmult}

\usepackage{mathptmx}       % selects Times Roman as basic font
\usepackage{helvet}         % selects Helvetica as sans-serif font
\usepackage{courier}        % selects Courier as typewriter font
\usepackage{type1cm}        % activate if the above 3 fonts are
\usepackage{makeidx}         % allows index generation
\usepackage{graphicx}        % standard LaTeX graphics tool
\usepackage{multicol}        % used for the two-column index
\usepackage[bottom]{footmisc}% places footnotes at page bottom

\usepackage{amsmath,amssymb}
\usepackage{stmaryrd}
\usepackage{url}

\makeindex             % used for the subject index
\def\ZZ{{\mathbb Z}_2 \times {\mathbb Z}_2}
\def\g{\mathfrak{g}}
\def\ga#1{\gamma_{#1}}
\def\del#1{\partial_{#1}} 
\def\hf{\frac{1}{2}}
\def\tP#1{\tilde{P}_{#1}}
\def\tG#1{\tilde{G}_{#1}}
\def\tX#1{\tilde{X}_{#1}}
\def\xp#1{X^P_{#1}}
\def\xg#1{X^G_{#1}}
\def\SDsum{\supset \hspace{-1em}\hspace{-0.2pt}+}

\begin{document}

% Please name your Latex file as lastname_session.tex
% where session should be included as in the following examples:
% - lastname_M2O.tex for oral talk in M2,
% - lastname_P5P.tex for poster in P5,
% - lastname_PL.tex for plenary talks

\title*{Generalized supersymmetry and L\'evy-Leblond equation}
\author{N. Aizawa, Z. Kuznetsova, H. Tanaka and F. Toppan}
\institute{N. Aizawa, H. Tanaka \at Osaka Prefecture University,Sakai, Osaka 599-8531 Japan, \email{aizawa@p.s.osakafu-u.ac.jp,s_h.tanaka@p.s.osakafu-u.ac.jp}
\and Z. Kuznetsova \at UFABC, Av. dos Estados 5001, Bangu, cep 09210-580, Santo Andr\'e (SP), Brazil, \email{zhanna.kuznetsova@ufabc.edu.br}
\and F. Toppan \at CBPF, Rua Dr. Xavier Sigaud 150, Urca, cep 22290-180, Rio de Janeiro (RJ), Brazil, \email{toppan@cbpf.br}}
\maketitle

\abstract*{The symmetries of the L\'evy-Leblond equation are investigated beyond the standard Lie framework. 
It is shown that the equation has two remarkable symmetries. 
One is given by the super Schr\"odinger algebra and the other is by a $\ZZ$ graded Lie algebra. 
The $\ZZ$ graded Lie algebra is achieved by transforming bosonic into fermionic operators in the super Schr\"odinger algebra 
and introducing second order differential operators as generators of symmetry. 
}

\abstract{The symmetries of the L\'evy-Leblond equation are investigated beyond the standard Lie framework. 
It is shown that the equation has two remarkable symmetries. 
One is given by the super Schr\"odinger algebra and the other one by a $\ZZ$ graded Lie algebra. 
The $\ZZ$ graded Lie algebra is achieved by transforming bosonic into fermionic operators in the super Schr\"odinger algebra 
and introducing second order differential operators as generators of symmetry. 
\\
\begin{flushright}
CBPF-NF-005/16
\end{flushright}
}

\section{Introduction}

 The purpose of the present work is to show that a $\ZZ $ graded Lie algebra is a symmetry of a simple equation  of physics, 
the  L\'evy-Leblond equation (LLE), which is a non-relativistic wave equation of a spin $1/2$ particle \cite{LevyLeblondE}. 
In the process to prove the $\ZZ $ symmetry we also show that LLE has a supersymmetry given by the $ {\cal N}=1 $ super Schr\"odinger algebra (see
\cite{DuHo} and references therein). 

$ \ZZ $ graded Lie algebras (introduced in  \cite{RitWy,RitWy2}, see also \cite{Scheu})  are natural generalizations of Lie superalgebras.
We present their definition: 
Let $ \g $ be a vector space over $ \mathbb C $ or $ \mathbb R$ with a $ \ZZ $ grading structure, namely $\g$ is the direct sum of four distinct subspaces labelled by an element of the $ \ZZ $ group:
\begin{equation}
 \g = \g_{(0,0)} + \g_{(0,1)} + \g_{(1,0)} + \g_{(1,1)}.
\end{equation}
For two elements $ \vec{a} = (a_1, a_2),\ \vec{b}= (b_1, b_2)  \in \ZZ,$  we define 
\begin{equation}
   \vec{a} + \vec{b} = (a_1+b_1, a_2+b_2) \  (\text{mod}(2, 2)), \quad \vec{a}\cdot \vec{b} = a_1 b_1 + a_2 b_2 
\end{equation}
% An element  of $ \g_{\vec{a}} $ is denoted by  $ X_{\vec{a}}. $   
%
\begin{definition} \label{DEF:Z2Z2}
 If $ \g $ admits a bilinear form $ \llbracket \ , \ \rrbracket : \g \times \g \to \g $ satisfying the following three relations, 
 then $ \g $ is called a $ \ZZ $ graded Lie algebra:
   \begin{enumerate}
      \item $ \llbracket \g_{\vec{a}}, \g_{\vec{b}} \rrbracket \subseteq \g_{\vec{a}+\vec{b}}, $
      \item $ \llbracket X_{\vec{a}}, X_{\vec{b}} \rrbracket = -(-1)^{\vec{a} \cdot \vec{b} } \llbracket X_{\vec{b}}, X_{\vec{a}} \rrbracket,  $
      \item $ \llbracket X_{\vec{a}}, \llbracket X_{\vec{b}}, X \rrbracket \rrbracket = 
              \llbracket \llbracket X_{\vec{a}}, X_{\vec{b}}, \rrbracket, X \rrbracket 
              + (-1)^{\vec{a} \cdot \vec{b}} \llbracket X_{\vec{b}}, \llbracket X_{\vec{a}}, X \rrbracket \rrbracket, $
   \end{enumerate}
   where  $ X_{\vec{a}} \in \g_{\vec{a}}. $ 
\end{definition}
Two sub superalgebras exist (they are $ \g_{(0,0)} + \g_{(0,1)} $ and $ \g_{(0,0)} + \g_{(1,0)}). $ 
This fact plays a crucial role when the symmetry of the LLE is identified with a $\ZZ$ graded Lie algebra.  

 In contrast to ordinary Lie algebras and superalgebras, the number of papers in the literature discussing physical applications of 
$\ZZ$ graded Lie algebras is limited \cite{JaYaWy,LukRitt,Tol,Vasi,Zhel}. 
The equation discussed in this work is both simple and fundamental. Even so, we naturally encountered this unusual algebraic structure. 
This would suggest that $ \ZZ $ graded Lie algebras are natural objects in the investigation of symmetries.

  The plan of this paper is as follows. 
In the next section we introduce the LLE and present its symmetries. We show that the LLE has a super Schr\"odinger symmetry. In \S \ref{SEC:3} the supersymmetry is enhanced to a $\ZZ$ graded Lie symmetry.   

%%%%%%%%%%%%%%%%%%%%%%%%%%%%%%%%%%%%%%%%%%%%%%%%%%%%%%%%%%%%%%%%%%%%%%%%%%%%%%%%%
%
\section{LLE and its (super)symmetries} 

  The LLE here considered is a non-relativistic wave equation  for a spin $1/2$ free particle in $3$D space. 
The wavefunction is a four-component spinor $ \psi(x) ={}^T(\varphi_1(x),\varphi_2(x))$ 
where $ \varphi_a $ is a $ SU(2)$ spinor and $ x = (t, x_1,  x_2, x_3). $ 
We use the following form of LLE \cite{FuZh}:
\begin{equation}
\Omega \psi(x) = 0, \quad 
        \Omega = -2i \alpha \del{t} + i  \gamma_j \del{x_j} + 2m \beta,
\end{equation}
where the sum over the repeated index $ j = 1, 2, 3 $ is understood;
$ \gamma_{\mu}, \alpha, \beta $ are $ 4 \times 4 $ Dirac $\gamma$-matrices defined by
\begin{equation}
         \{ \gamma_{\mu}, \gamma_{\nu} \} = 2 g_{\mu\nu}, \quad 
      (g_{\mu\nu}) = \text{diag}(+,-,-,-), \quad \mu, \nu = 0, 1, 2, 3
\end{equation}
and
\begin{equation}
       \alpha = \hf (\ga{0}+\ga{4}), \quad \beta = \hf(\ga{0}-\ga{4}), \quad \ga{4} = \ga{0} \ga{1} \ga{2} \ga{3}.
\end{equation}
One may take any four dimensional representation of the $ \gamma$-matrices. We do not distinguish upper and lower indices since we are working in a non-relativistic setting. 
LLE is the square root of the free Schr\"odinger equation, namely $ \Omega^2 $ gives the free particle Schr\"odinger operator:
\begin{equation}
 \Omega^2 = -4im \del{t} +  \partial x_j^2.
\end{equation}

  We introduce now the symmetries of LLE. According to \cite{FuZh} we define them  in terms of symmetry operators \cite{FuZh}:
\begin{definition}\label{DEF:SymOP}
 Let $ \cal A $ be an operator acting on the solution space of LLE. 
Namely, $ \cal A $ maps a solution of LLE into another one: 
  \begin{equation}
              \Omega \psi = 0 \quad \Longrightarrow  \quad \Omega ({\cal A} \psi) \Big|_{\Omega \psi=0} = 0
  \end{equation}
In this case $ \cal A $ is called a symmetry operator. 
\end{definition}
In this definition $ \cal A $ can be any kind of operator such as multiplication, differential, integral,  \textit{etc}. 
The traditional Lie point symmetry group of differential equations is generated by a subset of symmetry operators 
which is  closed under commutations.  
Similarly, if a subset of symmetry operators forms a superalgebra or a $ \ZZ $ graded Lie algebra, then 
the set generates a graded group of transformations in the solution space of LLE.

 We  restrict now $\cal A$ to a differential operator of finite order. 
In this case a sufficient condition of symmetry is given as follows.
If $ \cal A $ satisfies either the condition
\begin{equation}
           [\Omega, {\cal A}] = \Lambda_{\cal A}(x) \Omega \label{OnShellComm}
\end{equation}
or
\begin{equation}
           \{\Omega, {\cal A}\} = \Gamma_{\cal A}(x) \Omega \label{OnShellAnti},
\end{equation}
where $ \Lambda_{\cal A}(x) $ or $ \Gamma_{\cal A}(x) $ is a $ 4 \times 4 $ matrix depending on the spacetime 
coordinates,  
then $ \cal A $ is a symmetry operator.

  We are looking for symmetry operators given by a first order differential operator. 
The results are summarized in the following two propositions:
\begin{proposition} \label{Prop1}
 The operators below are LLE symmetry operators satisfying the condition (\ref{OnShellComm}):
 \begin{eqnarray}
   P_j &=& \del{x_j}, \qquad G_j = t \del{x_j} + 2im x_j + \alpha \ga{j}, \qquad M = 2im,
   \nonumber \\
    H &=& \del{t}, \qquad D = 2t \del{t} + x_j \del{x_j} + 2 - \hf \ga{0} \ga{4}, 
    \nonumber \\
    K &=& tD - t^2 \del{t} + im x_j x_j + \alpha x_j \ga{j}, 
    \nonumber \\
    J_{jk} &=& x_j \del{x_k} - x_k \del{x_j} - \hf \ga{j} \ga{k},
    \nonumber \\
     \tX{j} &=& -\epsilon_{jkn} \Big( [\alpha, \ga{k}]\del{x_n} + \frac{im}{2} [\ga{k},\ga{n}] \Big).
 \end{eqnarray}
 The only two non-vanishing $ \Lambda_{\cal A}(x)$ matrices are $\Lambda_{D} = 1$, $\Lambda_{K}=t. $  
 For convenience the $ 4 \times 4 $ unit matrix $ {\bf 1}_4 $ is not explicitly indicated (e.g. $ P_j = {\bf 1}_4\, \del{x_j}\equiv \partial_{x_j}).$ 
\end{proposition}
Apart from the $\tX{j}$'s, the remaining symmetry operators close a Lie algebra.  
$ \mathfrak{h}(3) = \langle\;  P_j, G_j, M \;\rangle $ is the three dimensional Heisenberg Lie algebra with $M$ as a central element. 
We have the non-relativistic conformal algebra 
$ \mathfrak{sl}(2,{\mathbb R}) = \langle\;  H, D, K \;\rangle \ $ 
and the spatial rotation 
$ \mathfrak{so}(3) = \langle\;  J_{jk} \;\rangle. $ 
Combining together these three Lie algebras we get the Schr\"odinger algebra, whose structure is given by 
$ ( \mathfrak{sl}(2,{\mathbb R}) \oplus \mathfrak{so}(3) )$ $ \SDsum \; $ $ \mathfrak{h}(3) $, with $ \SDsum $ a semidirect sum of Lie algebras. 
We thus see that the Schr\"odinger group is a symmetry of LLE. 
This fact is already known in the literature. In \cite{FuZh} the Schr\"odinger algebra is presented as the maximal Lie symmetry of LLE.
  If the symmetry operators $ \tX{j} $ are included we are no longer able to close a Lie algebra. Their addition leads to a $ \ZZ $ graded Lie algebra. 
Before addressing the $\ZZ$ structure we look at the LLE's supersymmetry.

\begin{proposition}
 The operators below are LLE symmetry operators satisfying the condition (\ref{OnShellAnti}):
 \begin{eqnarray}
       Q &=& \frac{1}{\sqrt{-im}} \alpha \del{t} + \sqrt{-im} \beta, 
        \nonumber  \\
       S &=& \frac{1}{\sqrt{-im}} \alpha \Big( t\del{t} + x_j \del{x_j} + \frac{3}{2} \Big) + \sqrt{-im} ( t\beta + x_j \ga{j}),
        \nonumber \\ 
       X_j &=& \frac{1}{\sqrt{-im}} \alpha \del{x_j} + \sqrt{-im} \ga{j},  
 \end{eqnarray}
 with only one non-vanishing $ \Gamma_{\cal A}(x)$ matrix given by $ \Gamma_{S} = -\alpha/\sqrt{-im}. $   
\end{proposition}
The physical meaning of these symmetry operators becomes clear when computing their anticommutators:
     \begin{eqnarray}
          & & \{Q, Q\} = 2H, \qquad \{S, S\} = 2K, \qquad \{X_j, X_k\} = \delta_{jk} M,
           \nonumber \\
          & & \{Q, S\} = D, \qquad\ \{Q, X_j\} = P_j, \qquad \{S, X_j\} = G_j.
     \end{eqnarray}
It follows that $Q, S$ are, respectively, a supercharge and a conformal supercharge, with $ X_j$ a fermionic counterpart 
of $ \mathfrak{h}(3).$ 
Indeed, the Schr\"odinger algebra of Proposition \ref{Prop1} 
and $ \langle \; Q, S, X_j \; \rangle $ close the $ {\cal N}=1 $ super Schr\"odinger algebra. 
This is verified by direct computation of the (anti)commutation relations. 
The operator $Q$ is already found in \cite{FuZh} without recognizing it as a supercharge.  
One may also show (we omit the proof for space reasons), that there exists no other supercharge $ \overline{Q} $ 
satisfying
\begin{eqnarray}
   \{\overline{Q}, \overline{Q}\} &=& 2H, \qquad \, \{ Q, \overline{Q} \} = 0,
   \qquad 
   \{\overline{Q}, \Omega \} = \Gamma_{\overline{Q}}(x) \Omega,
   \nonumber \\[2pt]
   [D, \overline{Q}] &=& -\overline{Q}, \qquad [J_{jk}, \overline{Q}] = 0.
   \label{SuperChargeConditions}
\end{eqnarray}
We thus have the theorem:
\begin{theorem}
 The $ {\cal N} = 1 $ super Schr\"odinger algebra generates a symmetry supergroup of LLE and $ {\cal N} = 1 $ is the maximal supersymmetry. 
\end{theorem}

The supersymmetry of LLE was conjectured many years ago in the study of the worldline supersymmetry of the spinning particle \cite{GaGoTo}. 
If the symmetry is defined according to Definition \ref{DEF:SymOP}, then the conjecture is true. 
We mention here two other previous works on supersymmetry of LLE. 
In \cite{Horv} it was shown that LLE coupled with an arbitrary static magnetic field has a super Schr\"odinger symmetry.  
In \cite{HoMiPl}  the Dirac equation and the Deser-Jackiw-Templeton equation in a $(2+1)$ dimensional spacetime are unified in 
a single multiplet of $\mathfrak{osp}(1|2). $  It is shown that the non-relativistic limit of this system carries an $ {\cal N}=2$ super 
Schr\"odinger symmetry.

%%%%%%%%%%%%%%%%%%%%%%%%%%%%%%%%%%%%%%%%%%%%%%%%%%%%%%%%%%%%%%%%%%%%%%%%%%%%%%%
%
\section{The $\ZZ$ graded symmetry of LLE}
\label{SEC:3}

  In this section we consider the symmetry of LLE with the $\tX{j}$ operators. There are two key 
observations:  
(i) the $ \tX{j} $'s are obtained from the commutators of the fermionic generators $ X_j$, 
$ \tX{j} = \hf \epsilon_{jkn} [X_k, X_n]; $  
(ii) each pair $ (Q,S), (P_j, G_j) $ is a $ sl(2,{\mathbb R}) $-doublet under the adjoint action. 
The observation (i) implies that we need to give up the super Schr\"odinger structure, 
while (ii) implies that we may regard $(P_j,G_j)$ as fermionic since this treats all $ sl(2,{\mathbb R})$ doublets 
on equal footing \cite{Top}. 
Therefore we introduce, from the anticommutators, the new operators 
\begin{eqnarray}
  \tP{jk} &=& \{ P_j, P_k \}, \quad \tG{jk} = \{G_j, G_k\}, \quad 
  W_{jk} = \{P_j, G_k\}, 
  \nonumber \\
  \xp{jk} &=& \{ P_j, X_k \}, \quad \xg{jk} = \{G_j, X_k\}.
\end{eqnarray}
They are second order differential operators; it is easy to verify that they are symmetry operators of LLE. 
Surprisingly, these second order operators, together with the first order operators in the super Schr\"odinger algebra, close a $ \ZZ $ graded Lie algebra ${\cal G}_{\ZZ}$. 
This means that their (anti)commutators never produce higher order differential operators. 
The assignment of the grading is given by
\begin{eqnarray}
   \g_{00} &=& \langle \; H,\, D,\, K,\, J_{jk},\, \tX{j}, \, W_{jk}, \, \tP{jk},\, \tG{jk} \; \rangle,
   \nonumber \\
   \g_{01} &=& \langle \; P_j, \, G_j \; \rangle,
   \nonumber \\
   \g_{10} &=& \langle \; Q, \, S, \, \xp{jk}, \, \xg{jk} \; \rangle,
   \nonumber \\
   \g_{11} &=& \langle \; X_j \; \rangle. \label{Z2Z2alg}
\end{eqnarray}
One may verify, by direct but cumbersome computation of the (anti)commutators, that the algebra (\ref{Z2Z2alg}) satisfies Definition \ref{DEF:Z2Z2}. 
We remark that the multiplication operator $M$ has dropped out from this $ \ZZ $ graded Lie algebra. 
\begin{theorem}
 The $\ZZ$ graded Lie algebra defined by the operators in (\ref{Z2Z2alg}) generates a symmetry group of LLE. 
\end{theorem}

  We have shown, in summary, that LLE has a $ {\cal N}=1$ super Schr\"odinger symmetry and a $ \ZZ $ graded symmetry given by (\ref{Z2Z2alg}). 
The super Schr\"odinger algebra is not a subalgebra of the $\ZZ $ graded algebra, although they share the same symmetry operators. 
As a continuation of the present work one may investigate symmetries of LLE with a potential, 
since it is known that Schr\"odinger equation with linear or quadratic potential has the same symmetry as the free equation \cite{Boy,Nie}. 
It is also an interesting problem to study symmetries of a LLE for an arbitrary space dimension. 
This would be done systematically by making use of the representation theory of Clifford algebra. 
These works are in progress. Part of these results are reported in \cite{AKTT}.

%%%%%%%%%%%%%%%%%%%%%%%%%%%%%%%%%%%%%%%%%%%%%%%%%%%%%%%%%%%%%%%%%%%%%%%%%%%%%%%

\begin{acknowledgement}
N. A. is supported by the  grants-in-aid from JSPS (Contract No. 26400209). F. T. received support from
CNPq (PQ grant No. 306333/2013-9).
\end{acknowledgement}
%
% \section*{Appendix}
% \addcontentsline{toc}{section}{Appendix}
%
%

%\input{referenc}

\end{document}